\documentclass[twocolumn,twocolappendix]{aastex631}
\usepackage{amsmath}
\usepackage{booktabs}

\begin{document}

\title{Stability of C$_{59}$ Knockout Fragments from Femtoseconds to Infinity}

\author[0000-0003-1028-7976]{Michael Gatchell}
\author[0009-0002-2416-3181]{Naemi Florin}
\author[0000-0001-7776-5448]{Suvasthika Indrajith}
\author[0000-0003-4392-9867]{Jos\'{e} Eduardo Navarro Navarrete}
\author[0000-0002-8627-1009]{Paul Martini}
\author[0000-0001-8184-4595]{MingChao Ji}
\author{Peter Reinhed}
\author[0009-0001-7448-0030]{Stefan Ros\'en}
\author[0000-0002-6638-0291]{Ansgar Simonsson}
\author[0000-0002-0815-0658]{Henrik Cederquist}
\author[0000-0002-8209-5095]{Henning T.\ Schmidt}
\author[0000-0002-2493-4161]{Henning Zettergren}
\affiliation{Department of Physics, Stockholm University, 106 91 Stockholm, Sweden}

\begin{abstract}

We have studied the stability of C$_{59}$ anions as a function of time, from their formation on femtosecond timescales to their stabilization on second timescales and beyond, using a combination of theory and experiments. The C$_{59}^-$ fragments were produced in collisions between C$_{60}$ fullerene anions and neutral helium gas at a velocity of 90 km/s (corresponding to a collision energy of 166\,eV in the center-of-mass frame). The fragments were then stored in a cryogenic ion-beam storage ring at the DESIREE facility where they were followed for up to one minute. Classical molecular dynamics simulations were used to determine the reaction cross section and the excitation energy distributions of the products formed in these collisions. We found that about 15 percent of the C$_{59}^-$ ions initially stored in the ring are intact after about 100 ms, and that this population then remains intact indefinitely. 
This means that C$_{60}$ fullerenes exposed to energetic atoms and ions, such as stellar winds and shock waves, will produce stable, highly reactive products, like C$_{59}$, that are fed into interstellar chemical reaction networks. 
\end{abstract}


\section{Introduction} \label{sec:intro}

Of the more than 300 unique molecular species hitherto identified in the ISM, the largest are the C$_{60}$ and C$_{70}$ fullerenes \citep{McGuire:2022aa,Cami:2010aa,Sellgren:2010aa}. Fullerenes were first discovered in 1985 and they immediately sparked interest in astrophysical contexts \citep{Kroto:1985aa,Kroto:1988aa}. Their presence in space was eventually confirmed in the 2010s, first by vibrational transitions from planetary \citep{Cami:2010aa} and reflection nebulae \citep{Sellgren:2010aa}, and later when C$_{60}^+$ was identified by electronic transitions in the near-IR as the first identified carrier of diffuse interstellar bands (DIBs) \citep{Campbell:2015aa,Kuhn:2016aa,Linnartz:2020aa}.

Mechanisms leading to the formation and evolution of fullerene molecules in astronomical environments have been the focus of numerous studies since even before their discovery in space \citep{Candian:2019aa}. These can broadly be placed in two categories: top-down pathways where large molecules are broken down into fullerene cages \citep{Berne:2012aa,Zhang:2013aa,Zhen:2014aa,Berne:2015aa,Bernal:2019aa,Herrero:2022aa}; and bottom-up processes where smaller reactants combine to form a larger product \citep{Maul:2006aa,Zettergren:2010aa,Dunk:2012aa,Krasnokutski:2016aa,Delaunay:2018aa,Meng:2023aa}. A compelling top-down process is the processing of large polycyclic aromatic hydrocarbons (PAHs), another family of molecules in the ISM \citep{Tielens:2008aa}, which has been demonstrated to be a viable pathway both in theoretical calculations \citep{Berne:2012aa} and experiments \citep{Zhen:2014aa}. A key step there is the introduction of curvature to the planar structures through the conversion of hexagonal rings to pentagonal rings in the carbon lattice \citep{Berne:2012aa}. Studies of bottom-up processes have shown that the condensation of carbon in circumstellar environments can result in both individual fullerene molecules, but also multi-layered fullerene onions \citep{Meng:2023aa}.

In all such formation pathways, and likewise for the further evolution of existing fullerenes in interstellar and circumstellar environments, energetic processing plays an important role. This processing can take on many forms, such as radiation fields from stars, both in the UV range and at longer wavelengths \citep{Greenberg:2000aa,Moore:2009aa}; energetic electrons and ions in shock-heated gases and plasmas \citep{Slavin:2004aa,Micelotta:2010ab,Chakraborty:2024aa}; or energetic atoms or ions in stellar winds and supernova shock waves \cite{Micelotta:2010ab,Micelotta:2010aa}. Each of these are capable of breaking down molecular material and are the driving forces behind the rich chemical makeup of the interstellar medium (ISM) \citep{Tielens:2013aa,Herbst:2017aa}. When molecules are excited by such processes they will cool by a combination of radiative cooling \citep{Andersen:2001aa}, thermionic emission of electrons \citep{Andersen:2001aa,Andersen:2002ab}, or by fragmentation \citep{Tomita:2001ab}. For fullerenes like C$_{60}$ the latter usually involves the emission of one or several C$_2$ units in a statistical process that favors the lowest energy dissociation pathway \citep{Rohmund:1996aa,Tomita:2001ab}. The products are smaller, even-numbered fullerenes such as C$_{58}$. One of the few exceptions to this occurs when fullerenes are impacted by atoms or ions at velocities ranging from a few tens to several hundreds of kilometers per second \citep{Larsen:1999aa,Tomita:2002aa,Gatchell:2014tk,Gatchell:2016aa}. Then, the incoming particles may directly remove a single C atom in a hard, head-on collision in a process know as knockout.

Knockout-driven fragmentation differs from statistical, thermally-driven dissociation processes in that energetically unfavorable pathways may occur with high probabilities. For C$_{60}$ this means that the loss of a single C atom (12--13\,eV dissociation energy \citep{Stockett:2018wq}) can become a competitive channel to C$_2$ loss (10\,eV \citep{Christian:1992wi,Tomita:2001ab}).

The knockout process was first identified with fullerenes in measurements performed at Aarhus University where C$_{60}$ anions were collided with rare gas targets at center-of-mass energies of a few hundred eV, resulting in the production of C$_{59}$ cations and anions in addition to the usual even-numbered fragments \citep{Larsen:1999aa,Tomita:2002aa}. Knockout processes were later reported with cationic C$_{60}$ molecular \citep{Gatchell:2014tk,Stockett:2018wq} and neutral [C$_{60}$]$_k$ cluster \citep{Zettergren:2013vp,Seitz:2013wc,PhysRevA.89.062708,Delaunay:2018aa} targets colliding with energetic atoms and ions, respectively. In the studies of clusters of fullerenes it was shown that the fragments produced by knockout are highly reactive and readily formed covalent bonds with neighboring molecules in the loosely bound clusters, leading to the efficient production of, e.g., dumbbell-like C$_{119}^+$ products \citep{Zettergren:2013vp,PhysRevA.89.062708}. For intact C$_{60}$ molecules, the kinetic energy threshold for fusing two colliding cages into a single molecule is approximately 60\,eV \citep{Campbell:2003tj}. In contrast, the reaction barrier for forming covalent bonds between a C$_{59}$ molecule and an intact C$_{60}$ neighbor is well below 1\,eV \citep{Zettergren:2013vp,PhysRevA.89.062708}. This dramatic difference readily demonstrates the high reactivity of C$_{59}$ in comparison to intact fullerenes.

A longstanding question since the knockout mechanism was identified is whether or not the fragments formed when an atom is knocked out of a molecule can survive in the gas phase on timescales longer than those of the typical experiment (microseconds). Knockout processes often induce significant heating of the damaged molecules. But if the fragments indeed are stable on extended timescales, then their high chemical reactivity will enable new reaction pathways which are far from being accessible for intact fullerenes. 

Knockout damaged fullerene molecules in interstellar and circumstellar environments could thus play a significant role in their chemical makeup. 
To achieve this, however, the fragments would need to stabilize (cool) by, e.g., the emission of electrons (typically microsecond timescales) or photons (up to millisecond timescales for vibronic de-excitation). 

A recent study involving the polycyclic aromatic hydrocarbon (PAH) molecule coronene \citep{Gatchell:2021tb} investigated the long-term stability issue using the cryogenic Double ElectroStatic Ion Ring ExpEriment (DESIREE) \citep{Thomas:2011aa,Schmidt:2013aa}. In those experiments, coronene cations were collided with He atoms at a center-of-mass energy of 105\,eV. Fragment ions that had lost a single C atom through knockout were stored in the storage ring and the number of ions remaining in the ring was monitored as a function of storage time. It was shown that a small fraction of the fragments were lost on millisecond timescales due to their high internal energy, but that approximately 80 percent of the stored fragments stabilized to remain intact indefinitely in isolation. The conclusion of that study was that knockout fragments such as those of coronene formed in C$_{24}$H$_{12}$ + He $\rightarrow$ C$_{23}$H$_x^+$ + C + $\ldots$ collisions may stabilize in the ISM \citep{Gatchell:2021tb}.

Here, we report on experiments where C$_{59}^-$ ions are produced in collisions between C$_{60}^-$ and He atoms at center-of-mass energies of 166\,eV and on the survival probability of these fragments on timescales ranging up to one minute in one of the DESIREE storage rings. We find that about 15 percent of the fragments formed in this way survive long enough to stabilize radiatively in the gas phase. This shows that odd-numbered fullerene radicals, such as C$_{59}$, can be produced and survive in astrophysical environments.

\begin{figure}[ht]
\center
\includegraphics[width=0.65\columnwidth]{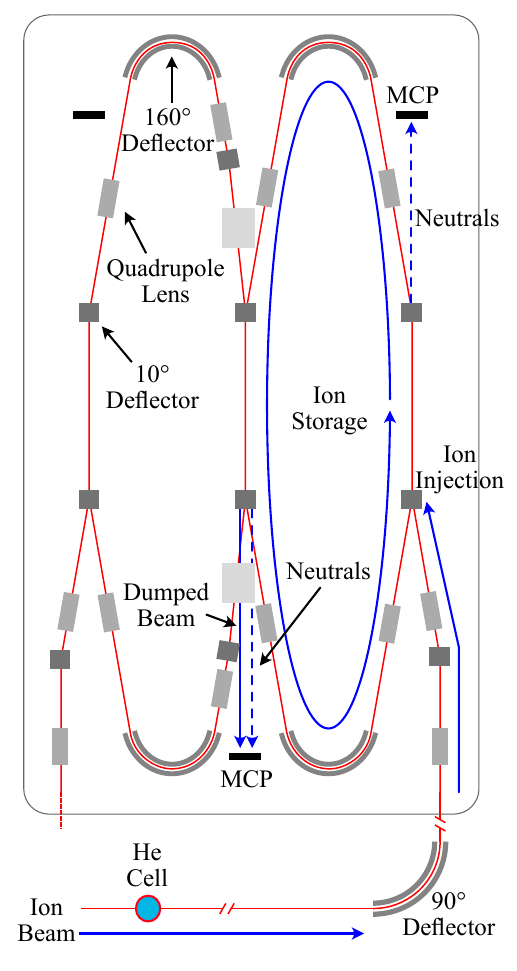} 
  \caption{Overview of the DESIREE storage rings. Mass-selected C$_{60}$ anions with kinetic energies of 30\,keV passed through a gas cell containing a dilute He gas in order to induce fragmentation in less than one percent of the fast C$_{60}^-$ ions. Selected beams of either C$_{60}^-$, C$_{59}^-$, or C$_{58}^-$ ions were stored in the rightmost DESIREE ion-beam storage ring for stability measurements. Electrostatic beamline elements are indicated by gray boxes and are labeled in the left ring. Neutral products formed along the two straight sections (dashed blue arrows) were detected by microchannel plate (MCP) detectors and dumped ions (solid blue line) were counted with the centrally located detector. The cryogenic region is indicated by the thin gray outline enclosing both rings.}
  \label{fig:DESIREE}
\end{figure} 

\section{Methods}

\subsection{Experimental methods}

The experiments were carried out at the DESIREE facility at Stockholm University \citep{Thomas:2011aa,Schmidt:2013aa}. The centerpiece of the DESIREE facility is a pair of cryogenically cooled electrostatic ion-beam storage rings with circumferences of approximately 8.7\,m each as outlined in Figure \ref{fig:DESIREE}. Here, we largely followed the experimental approach previously used to study the survival of PAH cations damaged by single-carbon knockout in energetic collisions \citep{Gatchell:2021tb} and to study the stability of C$_7^{2-}$ dianions \citep{Najeeb:2023aa}. Anions of C$_{60}$ were produced using a custom-made electron-attachment ion source based on the design by Yamada \emph{et al.}\ \citep{Yamada:2018aa}. We chose to study anions over cations to minimize the internal energies of the precursor ions and thus the influence of the ion source on the final results. The ions were accelerated by a 30\,kV potential before a beam chopper was used to produce ion pulses of suitable lengths for injection into the storage ring. A 102$^{\circ}$ bending magnet was used to select C$_{60}^-$ ions. These ions then passed through a 7\,cm long gas cell containing He gas at a pressure of $2\times 10^{-3}$\,mbar. The 30\,keV C$_{60}^-$ ions impacted the He target at a velocity of 90\,km/s, corresponding to a collision energy of 166\,eV in the center-of-mass reference frame.

Ions exiting the gas cell were analyzed and selected by scaling the voltages of all ion-optical elements after the gas cell with the kinetic energy of the anions, i.e.\ mass/initial mass corrected with a small energy loss in the gas cell. The ions (C$_{60}^-$, C$_{59}^-$, and C$_{58}^-$, respectively) were injected one species at a time and stored in one of the DESIREE storage rings for up to 100\,s, during which the spontaneous emission of neutral particles was monitored using the microchannel plate (MCP) detectors that follow each of the two straight sections. Any neutral particles formed along these straight sections will no longer be affected by the electrostatic deflectors that are used to steer the ion beam and will continue straight onto either detector as indicated by the dashed arrows in Fig.\ \ref{fig:DESIREE}.

We also performed measurements where the ions were dumped onto one of the detectors after a predetermined storage time. This was done by rapidly switching off the 10$^{\circ}$ deflector that follows the common straight section of the two ion storage rings (center of Figure \ref{fig:DESIREE}). This caused the ions circulating in the ring to continue straight onto the detector (the full arrow pointing towards one of the MCP detectors in Fig.\ \ref{fig:DESIREE}) that follows this section and allowed the number of ions in the ring to be counted as individual hits.

The two DESIREE storage rings share a common enclosure that is shown as the black outline in Figure \ref{fig:DESIREE}. This entire region is cryogenically cooled to a temperature of around 13\,K. The residual gas pressure in this chamber was on the order of $10^{-14}$\,mbar, which corresponds to a density of 10$^4$ particles (predominantly H$_2$) per cubic centimeter. These conditions allow ions to circulate in DESIREE over timescales ranging up to minutes or hours \citep{Backstrom:2015aa}, during which they are able to cool spontaneously with minimal environmental influence, e.g., from black-body radiation or collisions with residual gas particles \citep{Schmidt:2017aa}.

\subsection{Theoretical methods}

Following prior studies \citep{Stockett:2018wq}, simulations corresponding to the experiment were made using the LAMMPS molecular dynamics simulator, colliding single C$_{60}$ molecules with single He atoms. A non-periodic, cubic box with $40$\,Å sides was used, along with a time step of 0.1\,fs for the tracking of the motion of all individual atoms. The box was shrink-wrapped, meaning that its boundaries adjust in order to keep the atoms within it. A reactive empirical bond-order (REBO) potential acts between the C atoms, while the interaction between C and He is governed by a Ziegler-Biersack-Littmark (ZBL) potential. The former is an empirical, short ranged many-body potential for hydrocarbons \citep{Brenner:1990aa,Brenner:1992aa,Brenner:2002aa}, while the latter is a Coulombic potential developed to describe atomic high-energy collisions \citep{zbl_pot_book}. In contrast to the experiment, the simulations were made with a stationary C$_{60}$ molecule being hit by a He atom of 166\,eV (corresponding to the 30\,keV C$_{60}^-$ projectile hitting He atoms at rest in the lab frame).

\begin{figure}
\center
\includegraphics[width=0.75\columnwidth]{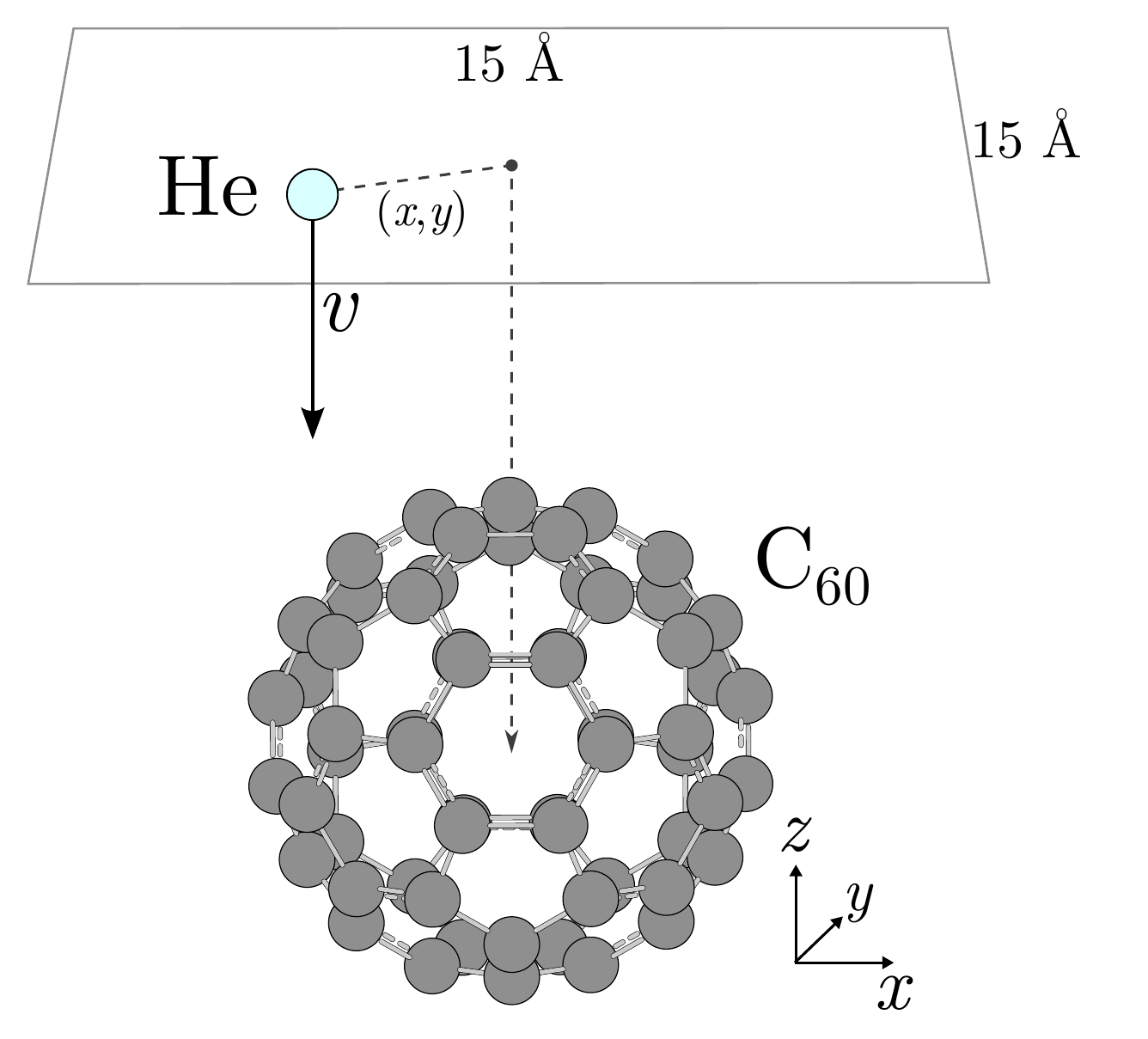} 
  \caption{Initial setup of a simulation. A helium atom is launched at a stationary and randomly oriented C$_{60}$ molecule with a velocity of 90 km/s. The fullerene is centered at the origin of a three-dimensional coordinate system. The initial position of the He atom is randomized in a square parallel to the $xy$-plane (c.f.\ text).}
  \label{fig:sim}
\end{figure} 

The C$_{60}$ structure, optimized with the REBO potential, was randomly rotated and positioned at the center of the simulation box. The projectile was then inserted with a fixed position in the $z$ direction and a randomized ($x,y$) position. The randomized rotation of the C$_{60}$ target relative to the randomized position of projectile counts for all possible initial positions and directions of the projectile in three-dimensional space. The initial position of the projectile was limited to a $15$\,Å $\times$ 15\,Å square in order to minimize the amount of runs where the projectile only transfers a negligible energy to the target (see Figure \ref{fig:sim}). In total, data was collected for 100,000 simulation runs.

In contrast to the experiments, these simulations are defined for neutral systems and do not take any electronic stopping into account. Fullerenes are large molecules with many delocalized valence electrons. Dissociation energies differ therefore only very slightly between neutral and anionic systems and in previous studies there has generally been a very good agreement between these types of simulations and experiments \citep{Stockett:2018wq}. Furthermore, in the present energy range, the energy transfer due to electronic stopping processes is negligible in comparison to those due to the nuclear scattering processes simulated here \citep{Gatchell:2016aa}. So, the exclusion of electronic stopping will not significantly influence the results of the simulations.

\section{Results and Discussion}

\subsection{DESIREE measurements}

\begin{figure}[t]
\center
\includegraphics[width=1\columnwidth]{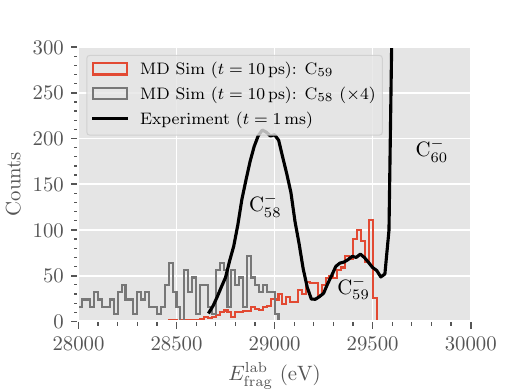} 
  \caption{Kinetic energy spectra of anionic fragments following collisions between  30\,keV C$_{60}^-$ and a He gas target. The solid black line shows the experimental spectrum and the histograms are from simulated collisions where components from each of the C$_{59}$ and C$_{58}$ (multiplied by four for clarity) products can be distinguished. The simulation data is shown in number of counts from 10$^5$ collisions with a bin width of 20\,eV. The normalization of the experimental data relative to the theoretical spectra is done arbitrarily.}
  \label{fig:ms}
\end{figure}

A kinetic energy spectrum of negatively charged products exiting the collision cell, measured using the 90$^{\circ}$ electrostatic deflector prior to injection into DESIREE, is shown as the solid black line in Figure \ref{fig:ms}. The energies shown here ($E^{\text{lab}}_{\text{frag}}$) are all given in the laboratory reference frame. The precursor C$_{60}^-$ ions carry a kinetic energy of 30\,keV with some inherent spread from the ion source and acceleration stage. In the gas cell, collisions with the He target introduce a recoil of varying degree causing an offset and broadening of the peaks relative to the expected positions based solely on mass loss. The small peak between 29.25 and 29.5 keV in Figure \ref{fig:ms} corresponds to C$_{59}^-$, where a single C atom has been knocked out, while the C$_{58}^-$ fragments are centered around 28.9\,keV. Only the wing of the C$_{60}^-$ peak is shown in the experimental spectrum since its maximum is orders of magnitude higher in intensity than those of the fragments. The intact C$_{60}^-$ ions mainly stem from the portion of the ion beam that passed through the gas cell without interacting with the dilute He target.

Corresponding spectra for C$_{58}$ and C$_{59}$ fragments from our MD simulations are also shown in Figure \ref{fig:ms}. Here, the changes in center-of-mass motion of the intact and damaged fullerene products caused by the collisions have been used to estimate the width and position of each peak. The position of the simulated C$_{59}$ distribution is consistent with the experimental data. This supports past observations that the knockout processes studied in the simulations are indeed responsible for the C$_{59}^-$ fragments detected in the experiments \citep{Larsen:1999aa,Stockett:2018wq,Gatchell:2014tk}.

\begin{figure}[t]
\center
\includegraphics[width=1\columnwidth]{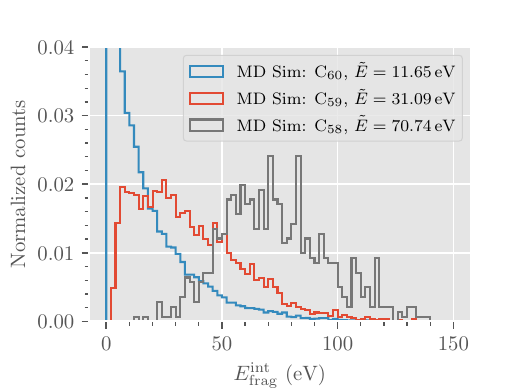} 
  \caption{Energy transfer distributions for C$_{60}$, C$_{59}$, and C$_{58}$ products determined from our MD simulations. For clarity, each distribution is normalized to a total integrated yield of unity. Each bin has a width of 2\,eV. Median energies of each distribution ($\Tilde{E}$) are given in the legend.}
  \label{fig:inte}
\end{figure}

At first glance, the agreement between experiments and theory is rather poor when it comes to the position and width of the C$_{58}^-$ peak. The reason for this is that, unlike in the experiments, \emph{every} C$_{58}$ fragment formed in the simulations originate from double knockout processes, collisions where a single He atom knocks out two C atoms from the fullerene cage. Under the conditions simulated here, the probability of this happening is smaller than that of single C knockout. This is a particularly violent process, with a large amount of energy being deposited into the cage. This can be seen in particular in Figure \ref{fig:inte} where the internal energy distribution of each fragment type formed in the simulation is shown. The high energy transfer also results in a significant broadening and shift of the C$_{58}$ spectrum in Figure \ref{fig:ms}, with the majority of the distribution falling outside of the range observed in the experiments. Because of this discrepancy between theory and experiments for the C$_{58}$ fragments, the bulk of these detected in the latter must therefore come from other processes than direct double knockout. A large portion of these are likely due to the later loss of a second C atom from the initially formed C$_{59}^-$ fragments in a subsequent statistical fragmentation process (discussed further in section \ref{sec:rates}). The loss of C$_{2}$ from C$_{60}^-$ ions excited in the collisions may also contribute to this peak.

The internal energy distributions ($E^{\text{int}}_{\text{frag}}$) for each of the main products, calculated from the MD simulations, are shown in Figure \ref{fig:inte}. As expected, the internal energy distribution of intact C$_{60}$ after the collisions is biased towards 0\,eV as many of the collisions, even with the box size used here, are at most glancing with little energy transfer. There is, however, a long tail extending past 100\,eV. The lowest dissociation energy of C$_{60}$ ions is about 10\,eV for C$_2$ loss \citep{Tomita:2001aa}, and about 40\,eV is required to activate these processes on microsecond timescales \citep{Qian:2013aa}. Furthermore, the electron affinity of C$_{60}$ is only 2.7\,eV \citep{Stochkel:2013aa,Huang:2014aa}. Thus, the hottest C$_{60}$ anions that survive the initial collisions with the He target will rapidly decay, either by electron emission or by fragmentation. 

The relaxation behavior of C$_{59}^-$ is to some extent similar to that of C$_{60}^-$. The knockout of a single C atom is potentially a destructive process, but a significant fraction of these trajectories result in a clean hit with a relatively small amount of energy transfer to the rest of the fullerene cage. The differences between the C$_{59}$ and C$_{60}$ distributions are thus relatively small considering the excess counts near 0\,eV in the latter case. In comparison, the aforementioned energy distribution for the C$_{58}$ fragments is significantly wider and peaks near 60\,eV. Here, only two simulation runs give C$_{58}$ double knockout with internal energies less than 20\,eV. Compared to similar studies with coronene, the energy transfer for the different products follow similar trends. The main difference is that the double knockout products here (C$_{58}$) display a stronger heating, relative to the other products, than was observed for the planar PAH and its fragments \citep{Gatchell:2021tb}. This is a result of the three-dimensional structure of the fullerene cage which increases the likelihood of multiple interactions between the projectile (He) and a single C$_{60}$ target molecule.

Due to the internal energies of the intact and damaged fullerenes leaving the collision cell, a portion of the ions stored in DESIREE will spontaneously decay. Neutral products from this decay, either dissociation or electron emission, are detected using one of the particle detectors situated after each of the straight sections (Fig.\ \ref{fig:DESIREE}). The neutral signal from these processes tends to follow a ${\sim1/t}$ behavior as a function of the storage time due to the broad internal energy distribution of the population of stored ions \citep{Hansen:2021aa}. The measured decay curves for stored C$_{60}^-$, C$_{59}^-$, and C$_{58}^-$ ions are shown in Figure \ref{fig:spont} of the Appendix as their interpretation is outside the scope of this work.

While the neutral particle yield from the stored ions provides insight into ongoing decay processes, it does not represent the behavior of the full ensemble of ions in the ring. Ions that cool by other means, e.g., photon emission, or that have a low enough internal energy to be stable for the timescale of the experiment, will not decay, making them invisible to the neutral particle measurements. To probe the actual number of ions in the storage ring at any given time we instead use a technique where one of the ten-degree deflectors is switched to ground on a microsecond timescale, which effectively dumps the entire content of the ring onto one of the particle detectors (lower center of Fig.\ \ref{fig:DESIREE}).

\begin{figure}[t]
\center
\includegraphics[width=1\columnwidth]{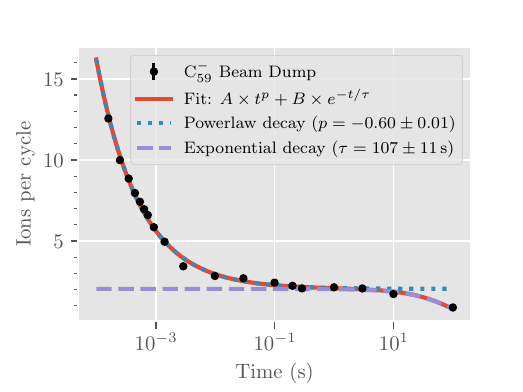} 
  \caption{Number of C$_{59}^-$ ions stored on average after different storage times, measured by dumping the contents of the ring onto a detector. Each point is the result of 500 measurement cycles (ion injection, storage, and beam dump). The solid line shows a fit to Eq.\ \ref{eq:fit}, with the two components shown as the dotted and dashed lines, respectively. Note the logarithmic horizontal scale that spans six orders of magnitude in time.}
  \label{fig:dump}
\end{figure} 

The signal of dumped C$_{59}^-$ ions, as function of time from their formation, is shown in Figure \ref{fig:dump} as the number of ions counted per measurement cycle. The shortest measurement achievable is when the ions were dumped after half of a turn in the ring, 160\,$\mu$s after they were produced. At the opposite end of the scale, the longest measurement involved storing the ions for 100\,s. It is immediately clear that the signal drops by about an order of magnitude during the first few hundred milliseconds before leveling out. At longer times, the ion-beam storage lifetime becomes a limiting factor, causing the signal to taper off once again. To quantify these processes we fit an expression of the type
 \begin{equation}
I(t) = A \times t^p + B \times e^{-t/\tau}
\label{eq:fit}
\end{equation}
to the data. Here, $t$ is the time since ion production (in the gas cell), while $A$, $B$, $p$, and $\tau$ are free parameters. The $p$ parameter is related to the rate at which a portion of the ions spontaneously decay due to their internal energies. The lifetime $\tau$ is the measured ion-beam storage lifetime that is dictated by the storage parameters and background conditions, and the parameter $B$ represents the expected ion signal in the absence of the storage-limited loss of ions. The fit of eq.\ \ref{eq:fit} is shown as the solid line in Figure \ref{fig:dump}. The key finding from this fit is the value of $B=2.05\pm0.02$. This value gives the average number of ions remaining in the ring when spontaneous loss of hot ions no longer occurs. This means that of the $12.6\pm 0.1$ C$_{59}^-$ ions that are counted after half of a turn in the ring for each injection (see Figure \ref{fig:dump}), $10.5\pm 0.2$ of them (about 85 percent) will dissociate or neutralize spontaneously on average. The two remaining ions, however, will remain intact indefinitely in isolation, i.e., in the gas phase. In our measurements, the number of remaining ions in the ring remains a constant for nearly three orders of magnitude in time, from approximately 100\,ms to 10\,s after the C$_{59}^-$ fragments are formed. After this the signal drops off due to imperfections in the storage conditions. The conditions at DESIREE allow for storage lifetimes on the order of hours \citep{Backstrom:2015aa,Kristiansson:2021aa}, but the small number of ions present in the experiment makes optimizing the ring parameters for longer lifetimes difficult. The $107\pm 11$\,s beam lifetime measured here thus represents an instrument optimization limitation and not a residual-gas limited ion-beam lifetime or a finite lifetime of the C$_{59}^-$ fragments. 

The results are in line with those found when coronene cation fragments were produced under similar experimental conditions \citep{Gatchell:2021tb}. The main difference is that there, as many as 80 percent of the fragments resulting from single C-loss stabilize during the first second of storage and will thus remain intact indefinitely. Here, the fraction is significantly smaller, with only 15 percent of the C$_{59}^-$ ions entering the ring remaining intact after this time. This difference is most likely due to the three-dimensional structure of the fullerene cages which leads to more energy being deposited on average in a collision.

\subsection{Destruction rate calculations}
\label{sec:rates}

To better understand these results and disentangle the different decay processes involved, we use statistical modeling to estimate the rates of the competing dissociation and electron detachment rates for our C$_{59}^-$ fragments. The rates of statistical decay processes such as these can be modeled using an Arrhenius rate equation 
\begin{equation}
k=\nu e^{-E_b/k_{\text{B}}T_{\text{eff}}}.
\label{eq:rates}
\end{equation}
Here, $k_{\text{B}}$ is the Boltzmann constant, $E_b$ is the activation (binding) energy for the reaction being modeled, $T_{\text{eff}}$ is the effective temperature of the system, and $\nu$ is a pre-exponential factor that dictates the rate of a specific process. \citet{Andersen:2001aa} showed that the relationship between temperature (for $T>1000$\,K) and internal energy, $E^{\text{int}}$, for an individual fullerene ion can be approximated by 
\begin{equation}
    T[\text{K}] = 1000 + (E^{\text{int}}[\text{eV}] - 7.4)/C,
\end{equation}
where $C$ is the heat capacity of the fullerene ion, which is assumed to be constant in the given energy range. The effective temperature, correcting for the finite size of the system \citep{Klots:1991aa}, is then given by $T_{\text{eff}}=T-E_b/2C$ \citep{Andersen:2001aa}. The heat capacity of C$_{59}$ is estimated by scaling the value for C$_{60}$ (0.0138\,eV/K \citep{Andersen:2001aa}) by the reduced number of internal degrees of freedom, 171 vs.\ 174, giving $C=0.0136$\,eV/K. The rates for different decay processes can then be obtained by inserting the corresponding values of $E_b$ and $\nu$ into the expressions above.

The electron affinity of C$_{59}$ is about 3.5\,eV, with the energy varying only slightly between the two main isomers \citep{PhysRevA.89.062708,Stockett:2018wq}. The isomers differ in which of the two ways the vacancy left after a C atom is removed from C$_{60}$ closes as the fragments relax, where either a pentagon-nonagon (pn) or heptagon-octagon (ho) pair or rings are formed \citep{PhysRevA.89.062708}. The pn isomer is energetically favored \citep{PhysRevA.89.062708}, but given their similar energetics we expect both to be present in our population of ions. In our calculations of the electron emission rate we therefore use $E_b^{\text{det}} = 3.5$\,eV, while the pre-exponential factor is taken to be the same as for electron detachment from C$_{60}^-$, $\nu^{\text{det}}=10^{13}$\,s$^{-1}$ \citep{Andersen:2000aa}.

In comparison, the lowest dissociation energy for C$_{59}^-$ is about 4.5\,eV for additional C-loss \citep{Stockett:2018wq}, which we use as our value for $E_b^{\text{dis}}$. The pre-exponential factor for this process is taken from \citet{Concina:2005aa} who determined the value for a range of fullerenes to be $\nu^{\text{dis}}=2\times10^{19}$\,s$^{-1}$.

\begin{figure}[h]
\center
\includegraphics[width=1\columnwidth]{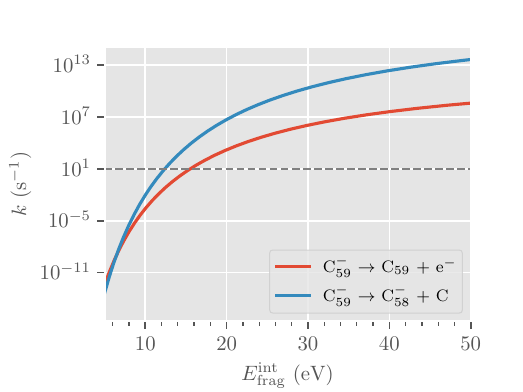} 
  \caption{Decay rates for electron detachment (red) and fragmentation through C-loss (blue) for C$_{59}^-$ fragments as function of their internal energies ($E_{\text{frag}}^{\text{int}}$). These were calculated using the Arrhenius rate equation (eq.\ (\ref{eq:rates})). The horizontal dashed line corresponds to the rate where quenching from radiative cooling begins to dominate, below which the fragments stabilize non-destructively.}
  \label{fig:rates}
\end{figure} 

The calculated rates for electron detachment (red) and fragmentation (blue) using this approach are shown in Figure \ref{fig:rates}. For all internal energies above 6.5\,eV, fragmentation is predicted to be the dominant destructive process for the population of C$_{59}^-$ fragments we store. The largest uncertainties in these calculations are the values of the pre-exponential factors. However, increasing or decreasing either of these by one or two orders of magnitude will only have a marginal effect on the trends in Figure \ref{fig:rates}. In Figure \ref{fig:dump} we can see that the ion decay is quenched after about 100\,ms as the C$_{59}^-$ fragments radiatively cool. This means that fragments with internal energies ($E_{\text{frag}}^{\text{int}}$) corresponding to decay rates where $k \lesssim 10^1$\,s$^{-1}$ will stabilize before they fragment or are neutralized. This cutoff corresponds to the dashed line in Figure \ref{fig:rates}. The crossing points between the this line and the curves for fragmentation and electron detachment rates occur at $E_{\text{frag}}^{\text{int}} \approx 12.5$\,eV and 15\,eV, respectively.

Our MD simulations (Figure \ref{fig:inte}) indicate that 16 percent of the C$_{59}^-$ fragments formed in our experiments have excitation energies below 12.5\,eV. This value is consistent with the fraction of fragments that survive the first second in our beam dump experiments, which was 15 percent of the initial population. The remaining fragments with energies greater than this will predominantly decay through C-loss according to the model. This is surprising at first as the the threshold for electron detachment from C$_{59}^-$ (3.5\,eV) is lower than that for C-loss (4.5\,eV), but is consistent with our interpretation of the mass spectrum in Figure \ref{fig:ms} in that the majority of the C$_{58}^-$ is formed from decaying knockout-damaged C$_{59}^-$. This also implies that the survival probabilities of C$_{59}$ fragments formed by knockout from neutral or cationic precursors should be similar to that for the anionic species, about 15 percent at the energies studied here, given their similar dissociation energies. We thus expect the charge state distributions of C$_{59}$ fragments in astronomical environments to mirror the local distributions for intact C$_{60}$.

\subsection{Reactivity of C$_{59}$ fragments}

\begin{figure}[h]
\center
\includegraphics[width=0.5\columnwidth]{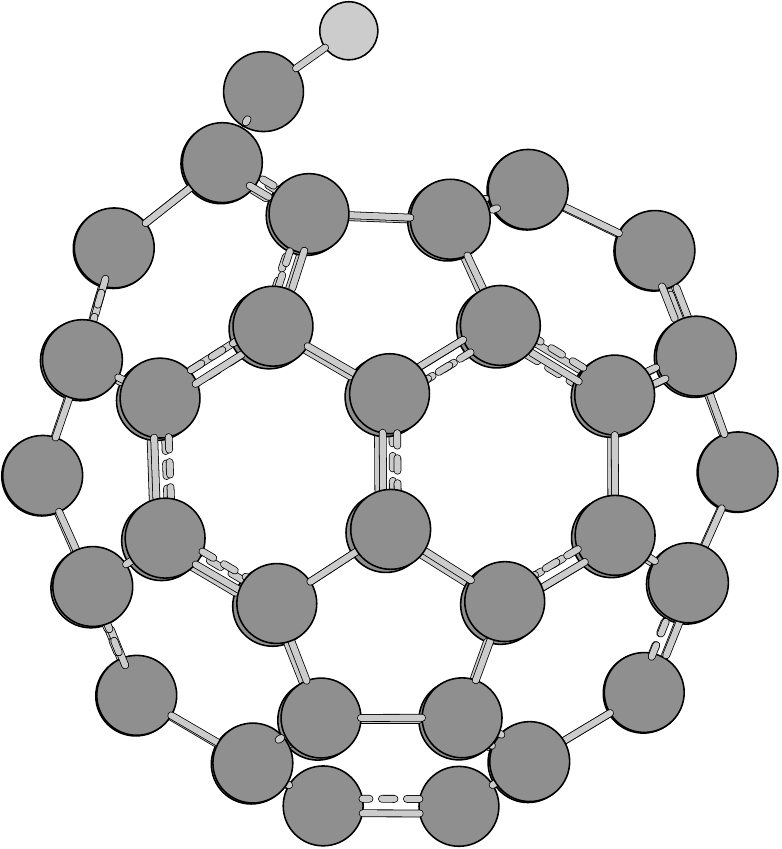} 
  \caption{Side view (perpendicular to the C-H bond) of the opmtimized structure of C$_{59}$H$^-$ calculated at B3LYP/6-31+G(d) level.}
  \label{fig:C59H}
\end{figure}

The C$_{59}$ fragments that do survive in the ISM can readily form covalent bonds with other species that they interact with. Previous work has demonstrated that the kinetic energy required to form a covalently bound dimer between a C$_{59}$ fragment and intact C$_{60}$ is less than 1\,eV, orders of magnitude lower than the energy required to fuse two pristine fullerenes \citep{PhysRevA.89.062708,Campbell:2003tj}. In most astronomical environments, interactions with atomic species like H or small molecular reactants with higher abundances than fullerenes are more probable. Hydrogenated fullerenes (fulleranes) have, for example, been proposed to be abundant in certain environments and a carrier of infrared and microwave emission bands from various sources \citep{Webster:1992aa,Iglesias-Groth:2006aa,Omont:2016aa,Zhang:2020aa}. Damaged C$_{59}$ molecules or ions could likewise form the basis for species of fulleranes with odd numbers of C atoms. Density functional theory (DFT) calculations presented in Table \ref{tab:BEs} show that the binding energy of H to C$_{59}$ is more than twice as high as that for H to C$_{60}$ for both neutrals and singly charged ions. An example of such a reaction product is C$_{59}$H$^-$ which is shown in Figure \ref{fig:C59H}. This increased reactivity is similar to that found for knockout damaged PAH molecules \citep{Chen:2014ab}. It is, however, localized to the position of the defect in the molecular cage, and once new bonds have formed there then the reactivity of the products should be similar to C$_{60}$ and other classical fullerenes.

\begin{table}[htbp]
   \begin{center}
       
    \caption{Binding energies (in eV) of H to C$_{59}$ and C$_{60}$ anions, neutrals, and cations, respectively, calculated at B3LYP/6-31+G(d) level using Gaussian 16 \citep{Frisch:2016aa}. The lowest energy structures for C$_{59}$H and C$_{59}$ are based on the pn isomer identified by \citet{PhysRevA.89.062708}.}

   \begin{tabular}{@{} lccc @{}} 

          & Anions  & Neutrals & Cations\\
      \toprule
      C$_{59}$H $\rightarrow$  C$_{59}$ + H (eV)     & 4.55 & 4.71  & 5.11\\
      C$_{60}$H $\rightarrow$  C$_{60}$ + H (eV)    & 2.16 & 1.83 & 2.49 \\
      \bottomrule
   \end{tabular}
  
   \label{tab:BEs}
   \end{center}
\end{table}

Overall, these findings suggest that exotic and reactive C$_{59}$ fragments, or products formed from them in subsequent reactions, most likely will be present and survive in regions where fullerenes are processed by energetic particles.

\section{Conclusions}

We have studied the stability of C$_{59}^-$ fragments formed in energetic collisions between C$_{60}$ anions and He atoms at a velocity of 90\,km/s. Despite the violent process in which they are formed, about 15 percent of the C$_{59}^-$ ions formed remain stable on timescales of seconds or longer, at which point they no longer decay. The ions lost during this time are expected to primarily decay through secondary dissociation (additional C-loss) on timescales up to about 100 milliseconds into C$_{58}^-$ fragments. The intact C$_{59}^-$ ions, however, are stabilized by radiative processes, e.g., vibrational de-excitation, and will remain intact indefinitely in isolation.

Astronomical fullerenes are expected to form mainly in the circumstellar regions of evolved stars from the outflow of carbon-rich material \citep{Goeres:1992aa,Cherchneff:2000aa,Jager:2009aa,Garcia-Hernandez:2012aa}. In these environments, wind-driven shockwaves play a major role in the processing and destruction of molecular material \cite{Rudnitskij:1997aa}. Likewise, energetic particles in supernova shockwaves and the heated post-shock gas can break down complex molecules \citep{Micelotta:2010ab,Micelotta:2010aa}. The velocities of particles in these regions are of the same order of magnitude as those studied in the present experiments. It is therefore likely that knockout processes occur and, as our results have shown, that odd-numbered fullerene radicals such as C$_{59}$ are present there as well and that their charge states mimic that of the C$_{60}$ precursors. Given their high reactivity, odd-numbered fullerene fragments are likely form new bonds with other atomic and molecular species, perhaps leading to reaction products unlikely to be produced by other means. Compared to C$_{60}$, the defect in the molecular cage introduced by the removal of a C atom should result in distinctive spectral fingerprints, both from vibrational transitions in infrared and electronic transitions at shorter wavelengths. Additional work will be needed, however, to provide reference spectra for the potential astronomical identification of these species or their reaction products.

This work was performed at the Swedish National Infrastructure, DESIREE (Swedish Research Council Contract Nos.\ 2017-00621 and 2021-00155). M.G., H.C., H.S., and H.Z.\ acknowledge support from the Swedish Research Council (contracts 2020-03104, 2019-04379, 2022-02822, and 2020-03437, respectively). It is a part of the project ``Probing charge- and mass-transfer reactions on the atomic level'', supported by the Knut and Alice Wallenberg Foundation (Grant no.\ 2018.0028). This article is based upon work from COST Actions CA18212 -- Molecular Dynamics in the GAS phase (MD-GAS) and CA21101 -- Confined Molecular Systems: From a New Generation of Materials to the Stars (COSY), supported by COST (European Cooperation in Science and Technology).

\begin{appendix}

\begin{figure}[h]
\center
\includegraphics[width=1\columnwidth]{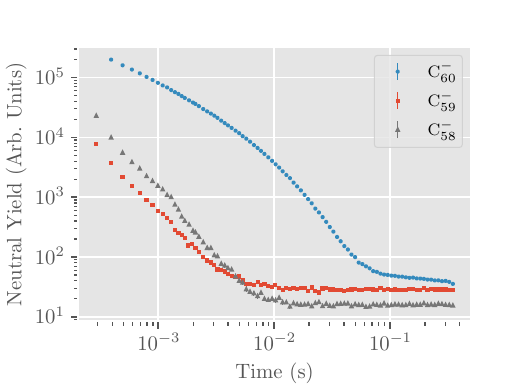} 
  \caption{The neutral particle yields from spontaneously decaying C$_{60}^-$ (blue), C$_{59}^-$ (red), and C$_{58}^-$ (gray) ions in DESIREE. The constant yields seen after 10\,ms with the C$_{59}^-$ and C$_{58}^-$ beams correspond to the detector background levels.}
  \label{fig:spont}
\end{figure}

Figure \ref{fig:spont} shows the neutral particle yield measured as a function of storage time for beams of C$_{60}^-$ (blue), C$_{59}^-$ (red), and C$_{58}^-$ (gray) in DESIREE.

\end{appendix}

\bibliography{library}{}

\begin{thebibliography}{}
\expandafter\ifx\csname natexlab\endcsname\relax\def\natexlab#1{#1}\fi
\providecommand{\url}[1]{\href{#1}{#1}}
\providecommand{\dodoi}[1]{doi:~\href{http://doi.org/#1}{\nolinkurl{#1}}}
\providecommand{\doeprint}[1]{\href{http://ascl.net/#1}{\nolinkurl{http://ascl.net/#1}}}
\providecommand{\doarXiv}[1]{\href{https://arxiv.org/abs/#1}{\nolinkurl{https://arxiv.org/abs/#1}}}

\bibitem[{Andersen \& Bonderup(2000)}]{Andersen:2000aa}
Andersen, J.~U., \& Bonderup, E. 2000, The European Physical Journal D -
  Atomic, Molecular, Optical and Plasma Physics, 11, 413,
  \dodoi{10.1007/s100530070070}

\bibitem[{Andersen {et~al.}(2002)Andersen, Bonderup, \&
  Hansen}]{Andersen:2002ab}
Andersen, J.~U., Bonderup, E., \& Hansen, K. 2002, Journal of Physics B:
  Atomic, Molecular and Optical Physics, 35, R1.
\newblock \url{http://stacks.iop.org/0953-4075/35/i=5/a=201}

\bibitem[{Andersen {et~al.}(2001)Andersen, Gottrup, Hansen, Hvelplund, \&
  Larsson}]{Andersen:2001aa}
Andersen, J.~U., Gottrup, C., Hansen, K., Hvelplund, P., \& Larsson, M.~O.
  2001, The European Physical Journal D - Atomic, Molecular, Optical and Plasma
  Physics, 17, 189, \dodoi{10.1007/s100530170022}

\bibitem[{B\"ackstr\"om {et~al.}(2015)B\"ackstr\"om, Hanstorp, Hole, Kaminska,
  Nascimento, Blom, Bj\"orkhage, K\"allberg, L\"ofgren, Reinhed, Ros\'en,
  Simonsson, Thomas, Mannervik, Schmidt, \& Cederquist}]{Backstrom:2015aa}
B\"ackstr\"om, E., Hanstorp, D., Hole, O.~M., {et~al.} 2015, Physical Review
  Letters, 114, 143003, \dodoi{10.1103/PhysRevLett.114.143003}

\bibitem[{Bernal {et~al.}(2019)Bernal, Haenecour, Howe, Zega, Amari, \&
  Ziurys}]{Bernal:2019aa}
Bernal, J.~J., Haenecour, P., Howe, J., {et~al.} 2019, The Astrophysical
  Journal, 883, L43, \dodoi{10.3847/2041-8213/ab4206}

\bibitem[{Bern{\'e} {et~al.}(2015)Bern{\'e}, Montillaud, \&
  Joblin}]{Berne:2015aa}
Bern{\'e}, O., Montillaud, J., \& Joblin, C. 2015, {Astronomy and
  Astrophysics}, 577, A133, \dodoi{10.1051/0004-6361/201425338}

\bibitem[{Bern{\'e} \& Tielens(2012)}]{Berne:2012aa}
Bern{\'e}, O., \& Tielens, A. G. G.~M. 2012, Proceedings of the National
  Academy of Sciences, 109, 401, \dodoi{10.1073/pnas.1114207108}

\bibitem[{Brenner(1990)}]{Brenner:1990aa}
Brenner, D.~W. 1990, Physical Review B, 42, 9458,
  \dodoi{10.1103/PhysRevB.42.9458}

\bibitem[{Brenner(1992)}]{Brenner:1992aa}
---. 1992, Physical Review B, 46, 1948, \dodoi{10.1103/PhysRevB.46.1948.2}

\bibitem[{Brenner {et~al.}(2002)Brenner, Shenderova, Harrison, Stuart, Ni, \&
  Sinnott}]{Brenner:2002aa}
Brenner, D.~W., Shenderova, O.~A., Harrison, J.~A., {et~al.} 2002, Journal of
  Physics: Condensed Matter, 14, 783

\bibitem[{Cami {et~al.}(2010)Cami, Bernard-Salas, Peeters, \&
  Malek}]{Cami:2010aa}
Cami, J., Bernard-Salas, J., Peeters, E., \& Malek, S.~E. 2010, Science, 329,
  1180, \dodoi{10.1126/science.1192035}

\bibitem[{Campbell(2003)}]{Campbell:2003tj}
Campbell, E. E.~B. 2003, Fullerene-Fullerene Collisions, ed. E.~E.~B. Campbell
  (Dordrecht: Springer Netherlands), 161--189,
  \dodoi{10.1007/1-4020-2524-6{\_}9}

\bibitem[{Campbell {et~al.}(2015)Campbell, Holz, Gerlich, \&
  Maier}]{Campbell:2015aa}
Campbell, E.~K., Holz, M., Gerlich, D., \& Maier, J.~P. 2015, Nature, 523, 322,
  \dodoi{10.1038/nature14566}

\bibitem[{Candian(2019)}]{Candian:2019aa}
Candian, A. 2019, Nature, 574, 490, \dodoi{10.1038/d41586-019-03139-6}

\bibitem[{Chakraborty {et~al.}(2024)Chakraborty, Yurchenko, Georges, Simon,
  Lacinbala, Chandrasekaran, Jayaram, Dartois, Kassi, Gusdorf, Lesaffre,
  Jagadeesh, Arunan, \& Biennier}]{Chakraborty:2024aa}
Chakraborty, S., Yurchenko, S.~N., Georges, R., {et~al.} 2024, A\&A, 681.
\newblock \url{https://doi.org/10.1051/0004-6361/202347035}

\bibitem[{Chen {et~al.}(2014)Chen, Gatchell, Stockett, Alexander, Zhang,
  Rousseau, Domaracka, Maclot, Delaunay, Adoui, Huber, Schlath{\"o}lter,
  Schmidt, Cederquist, \& Zettergren}]{Chen:2014ab}
Chen, T., Gatchell, M., Stockett, M.~H., {et~al.} 2014, The Journal of Chemical
  Physics, 140, 224306, \dodoi{http://dx.doi.org/10.1063/1.4881603}

\bibitem[{{Cherchneff} {et~al.}(2000){Cherchneff}, {Le Teuff}, {Williams}, \&
  {Tielens}}]{Cherchneff:2000aa}
{Cherchneff}, I., {Le Teuff}, Y.~H., {Williams}, P.~M., \& {Tielens},
  A.~G.~G.~M. 2000, {Astronomy and Astrophysics}, 357, 572

\bibitem[{Christian {et~al.}(1992)Christian, Wan, \&
  Anderson}]{Christian:1992wi}
Christian, J.~F., Wan, Z., \& Anderson, S.~L. 1992, The Journal of Physical
  Chemistry, 96, 10597

\bibitem[{Concina {et~al.}(2005)Concina, Tomita, Andersen, \&
  Hvelplund}]{Concina:2005aa}
Concina, B., Tomita, S., Andersen, J.~U., \& Hvelplund, P. 2005, The European
  Physical Journal D - Atomic, Molecular, Optical and Plasma Physics, 34, 191,
  \dodoi{10.1140/epjd/e2005-00142-5}

\bibitem[{Delaunay {et~al.}(2018)Delaunay, Gatchell, Mika, Domaracka, Adoui,
  Zettergren, Cederquist, Rousseau, \& Huber}]{Delaunay:2018aa}
Delaunay, R., Gatchell, M., Mika, A., {et~al.} 2018, Carbon, 129, 766,
  \dodoi{10.1016/j.carbon.2017.12.079}

\bibitem[{Dunk {et~al.}(2012)Dunk, Kaiser, Hendrickson, Quinn, Ewels,
  Nakanishi, Sasaki, Shinohara, Marshall, \& Kroto}]{Dunk:2012aa}
Dunk, P.~W., Kaiser, N.~K., Hendrickson, C.~L., {et~al.} 2012, Nature
  Communications, 3, 855.
\newblock \url{http://dx.doi.org/10.1038/ncomms1853}

\bibitem[{Frisch {et~al.}(2016)Frisch, Trucks, Schlegel, Scuseria, Robb,
  Cheeseman, Scalmani, Barone, Petersson, Nakatsuji, Li, Caricato, Marenich,
  Bloino, Janesko, Gomperts, Mennucci, Hratchian, Ortiz, Izmaylov, Sonnenberg,
  Williams-Young, Ding, Lipparini, Egidi, Goings, Peng, Petrone, Henderson,
  Ranasinghe, Zakrzewski, Gao, Rega, Zheng, Liang, Hada, Ehara, Toyota, Fukuda,
  Hasegawa, Ishida, Nakajima, Honda, Kitao, Nakai, Vreven, Throssell,
  Montgomery~Jr., Peralta, Ogliaro, Bearpark, Heyd, Brothers, Kudin,
  Staroverov, Keith, Kobayashi, Normand, Raghavachari, Rendell, Burant,
  Iyengar, Tomasi, Cossi, Millam, Klene, Adamo, Cammi, Ochterski, Martin,
  Morokuma, Farkas, Foresman, \& Fox}]{Frisch:2016aa}
Frisch, M.~J., Trucks, G.~W., Schlegel, H.~B., {et~al.} 2016, Gaussian 16 Rev.
  C.03

\bibitem[{Garc{\'\i}a-Hern{\'a}ndez {et~al.}(2012)Garc{\'\i}a-Hern{\'a}ndez,
  Villaver, Garc{\'\i}a-Lario, Acosta-Pulido, Manchado, Stanghellini, Shaw, \&
  Cataldo}]{Garcia-Hernandez:2012aa}
Garc{\'\i}a-Hern{\'a}ndez, D.~A., Villaver, E., Garc{\'\i}a-Lario, P., {et~al.}
  2012, The Astrophysical Journal, 760, 107

\bibitem[{Gatchell \& Zettergren(2016)}]{Gatchell:2016aa}
Gatchell, M., \& Zettergren, H. 2016, Journal of Physics B: Atomic, Molecular
  and Optical Physics, 49, 162001, \dodoi{10.1088/0953-4075/49/16/162001}

\bibitem[{Gatchell {et~al.}(2014)Gatchell, Stockett, Rousseau, Chen, Kulyk,
  Schmidt, Chesnel, Domaracka, M{\'e}ry, Maclot, Adoui, St{\o}chkel, Hvelplund,
  Wang, Alcam{\'\i}, Huber, Mart{\'\i}n, Zettergren, \&
  Cederquist}]{Gatchell:2014tk}
Gatchell, M., Stockett, M., Rousseau, P., {et~al.} 2014, International Journal
  of Mass Spectrometry, 365--366, 260 , \dodoi{10.1016/j.ijms.2013.12.013}

\bibitem[{Gatchell {et~al.}(2021)Gatchell, Ameixa, Ji, Stockett, Simonsson,
  Denifl, Cederquist, Schmidt, \& Zettergren}]{Gatchell:2021tb}
Gatchell, M., Ameixa, J., Ji, M., {et~al.} 2021, Nature Communications, 12,
  6646, \dodoi{10.1038/s41467-021-26899-0}

\bibitem[{{Goeres} \& {Sedlmayr}(1992)}]{Goeres:1992aa}
{Goeres}, A., \& {Sedlmayr}, E. 1992, \aap, 265, 216

\bibitem[{Greenberg {et~al.}(2000)Greenberg, Gillette, Mu{\~n}oz~Caro, Mahajan,
  Zare, Li, Schutte, de~Groot, \& Mendoza-G{\'o}mez}]{Greenberg:2000aa}
Greenberg, J.~M., Gillette, J.~S., Mu{\~n}oz~Caro, G.~M., {et~al.} 2000, The
  Astrophysical Journal, 531, L71, \dodoi{10.1086/312526}

\bibitem[{Hansen(2021)}]{Hansen:2021aa}
Hansen, K. 2021, Mass Spectrometry Reviews, 40, 725,
  \dodoi{https://doi.org/10.1002/mas.21630}

\bibitem[{Herbst(2017)}]{Herbst:2017aa}
Herbst, E. 2017, International Reviews in Physical Chemistry, 36, 287,
  \dodoi{10.1080/0144235X.2017.1293974}

\bibitem[{Herrero {et~al.}(2022)Herrero, Jim{\'e}nez-Redondo, Pel{\'a}ez,
  Mat{\'e}, \& Tanarro}]{Herrero:2022aa}
Herrero, V.~J., Jim{\'e}nez-Redondo, M., Pel{\'a}ez, R.~J., Mat{\'e}, B., \&
  Tanarro, I. 2022, Frontiers in Astronomy and Space Sciences, 9.
\newblock \url{https://www.frontiersin.org/articles/10.3389/fspas.2022.1083288}

\bibitem[{Huang {et~al.}(2014)Huang, Dau, Liu, \& Wang}]{Huang:2014aa}
Huang, D.-L., Dau, P.~D., Liu, H.-T., \& Wang, L.-S. 2014, The Journal of
  Chemical Physics, 140, 224315, \dodoi{10.1063/1.4881421}

\bibitem[{Iglesias-Groth(2006)}]{Iglesias-Groth:2006aa}
Iglesias-Groth, S. 2006, Monthly Notices of the Royal Astronomical Society,
  368, 1925, \dodoi{10.1111/j.1365-2966.2006.10272.x}

\bibitem[{{J{\"a}ger} {et~al.}(2009){J{\"a}ger}, {Huisken}, {Mutschke},
  {Jansa}, \& {Henning}}]{Jager:2009aa}
{J{\"a}ger}, C., {Huisken}, F., {Mutschke}, H., {Jansa}, I.~L., \& {Henning},
  T. 2009, The Astrophysical Journal, 696, 706,
  \dodoi{10.1088/0004-637X/696/1/706}

\bibitem[{Klots(1991)}]{Klots:1991aa}
Klots, C.~E. 1991, Zeitschrift f{\"u}r Physik D Atoms, Molecules and Clusters,
  21, 335, \dodoi{10.1007/BF01438406}

\bibitem[{Krasnokutski {et~al.}(2016)Krasnokutski, Kuhn, Kaiser, Mauracher,
  Renzler, Bohme, \& Scheier}]{Krasnokutski:2016aa}
Krasnokutski, S.~A., Kuhn, M., Kaiser, A., {et~al.} 2016, The Journal of
  Physical Chemistry Letters, 7, 1440, \dodoi{10.1021/acs.jpclett.6b00462}

\bibitem[{Kristiansson {et~al.}(2021)Kristiansson, Schiffmann, Grumer, Karls,
  de~Ruette, Eklund, Ideb{\"o}hn, Gibson, Brage, Zettergren, Hanstorp, \&
  Schmidt}]{Kristiansson:2021aa}
Kristiansson, M.~K., Schiffmann, S., Grumer, J., {et~al.} 2021, Physical Review
  A, 103, 062806, \dodoi{10.1103/PhysRevA.103.062806}

\bibitem[{Kroto(1988)}]{Kroto:1988aa}
Kroto, H. 1988, Science, 242, 1139, \dodoi{10.1126/science.242.4882.1139}

\bibitem[{Kroto {et~al.}(1985)Kroto, Heath, O'Brien, Curl, \&
  Smalley}]{Kroto:1985aa}
Kroto, H.~W., Heath, J.~R., O'Brien, S.~C., Curl, R.~F., \& Smalley, R.~E.
  1985, Nature, 318, 162

\bibitem[{Kuhn {et~al.}(2016)Kuhn, Renzler, Postler, Ralser, Spieler, Simpson,
  Linnartz, Tielens, Cami, Mauracher, Wang, Alcam{\'\i}, Mart{\'\i}n, Beyer,
  Wester, Lindinger, \& Scheier}]{Kuhn:2016aa}
Kuhn, M., Renzler, M., Postler, J., {et~al.} 2016, Nature Communications, 7,
  13550

\bibitem[{Larsen {et~al.}(1999)Larsen, Hvelplund, Larsson, \&
  Shen}]{Larsen:1999aa}
Larsen, M., Hvelplund, P., Larsson, M., \& Shen, H. 1999, The European Physical
  Journal D - Atomic, Molecular, Optical and Plasma Physics, 5, 283,
  \dodoi{10.1007/s100530050257}

\bibitem[{Linnartz {et~al.}(2020)Linnartz, Cami, Cordiner, Cox, Ehrenfreund,
  Foing, Gatchell, \& Scheier}]{Linnartz:2020aa}
Linnartz, H., Cami, J., Cordiner, M., {et~al.} 2020, Journal of Molecular
  Spectroscopy, 367, 111243, \dodoi{10.1016/j.jms.2019.111243}

\bibitem[{Maul {et~al.}(2006)Maul, Berg, Marosits, Sch\"onhense, \&
  Huber}]{Maul:2006aa}
Maul, J., Berg, T., Marosits, E., Sch\"onhense, G., \& Huber, G. 2006, Physical
  Review B, 74, 161406, \dodoi{10.1103/PhysRevB.74.161406}

\bibitem[{McGuire(2022)}]{McGuire:2022aa}
McGuire, B.~A. 2022, The Astrophysical Journal Supplement Series, 259, 30,
  \dodoi{10.3847/1538-4365/ac2a48}

\bibitem[{Meng \& Wang(2023)}]{Meng:2023aa}
Meng, Z., \& Wang, Z. 2023, Monthly Notices of the Royal Astronomical Society,
  526, 3335, \dodoi{10.1093/mnras/stad2754}

\bibitem[{Micelotta {et~al.}(2010{\natexlab{a}})Micelotta, Jones, \&
  Tielens}]{Micelotta:2010ab}
Micelotta, E.~R., Jones, A.~P., \& Tielens, A. G. G.~M. 2010{\natexlab{a}},
  {Astronomy and Astrophysics}, 510, A37, \dodoi{10.1051/0004-6361/200911683}

\bibitem[{Micelotta {et~al.}(2010{\natexlab{b}})Micelotta, Jones, \&
  Tielens}]{Micelotta:2010aa}
---. 2010{\natexlab{b}}, Astronomy and Astrophysics, 510, A36,
  \dodoi{10.1051/0004-6361/200911682}

\bibitem[{Moore \& Hudson(2009)}]{Moore:2009aa}
Moore, M.~H., \& Hudson, R.~L. 2009, Bulletin of the American Astronomical
  Society, 41, 717.
\newblock \url{https://ui.adsabs.harvard.edu/abs/2009AAS...21421202M}

\bibitem[{Najeeb {et~al.}(2023)Najeeb, Stockett, Anderson, Kristiansson,
  Reinhed, Simonsson, Ros\'en, Thomas, Chartkunchand, Gnaser, Golser, Hanstorp,
  Larson, Cederquist, Schmidt, \& Zettergren}]{Najeeb:2023aa}
Najeeb, P.~K., Stockett, M.~H., Anderson, E.~K., {et~al.} 2023, Phys. Rev.
  Lett., 131, 113003, \dodoi{10.1103/PhysRevLett.131.113003}

\bibitem[{Omont(2016)}]{Omont:2016aa}
Omont, A. 2016, A\&A, 590.
\newblock \url{https://doi.org/10.1051/0004-6361/201527685}

\bibitem[{Qian {et~al.}(2013)Qian, Ma, Chen, Li, Zhu, Zhang, Martin, Br\'edy,
  Bernard, \& Chen}]{Qian:2013aa}
Qian, D.~B., Ma, X., Chen, Z., {et~al.} 2013, Phys. Rev. A, 87, 063201,
  \dodoi{10.1103/PhysRevA.87.063201}

\bibitem[{Rohmund {et~al.}(1996)Rohmund, Glotov, Hansen, \&
  Campbell}]{Rohmund:1996aa}
Rohmund, F., Glotov, A.~V., Hansen, K., \& Campbell, E. E.~B. 1996, Journal of
  Physics B: Atomic, Molecular and Optical Physics, 29, 5143.
\newblock \url{http://stacks.iop.org/0953-4075/29/i=21/a=025}

\bibitem[{Rudnitskij(1997)}]{Rudnitskij:1997aa}
Rudnitskij, G.~M. 1997, Astrophysics and Space Science, 251, 259,
  \dodoi{10.1023/A:1000719008354}

\bibitem[{Schmidt {et~al.}(2013)Schmidt, Thomas, Gatchell, Ros{\'e}n, Reinhed,
  L{\"o}fgren, Br{\"a}nnholm, Blom, Bj{\"o}rkhage, B{\"a}ckstr{\"o}m,
  Alexander, Leontein, Hanstorp, Zettergren, Liljeby, K{\"a}llberg, Simonsson,
  Hellberg, Mannervik, Larsson, Geppert, Rensfelt, Danared, Pa{\'a}l, Masuda,
  Halld{\'e}n, Andler, Stockett, Chen, K{\"a}llersj{\"o}, Weimer, Hansen,
  Hartman, \& Cederquist}]{Schmidt:2013aa}
Schmidt, H.~T., Thomas, R.~D., Gatchell, M., {et~al.} 2013, Review of
  Scientific Instruments, 84, 055115, \dodoi{10.1063/1.4807702}

\bibitem[{Schmidt {et~al.}(2017)Schmidt, Eklund, Chartkunchand, Anderson,
  Kami\ifmmode~\acute{n}\else \'{n}\fi{}ska, de~Ruette, Thomas, Kristiansson,
  Gatchell, Reinhed, Ros\'en, Simonsson, K\"allberg, L\"ofgren, Mannervik,
  Zettergren, \& Cederquist}]{Schmidt:2017aa}
Schmidt, H.~T., Eklund, G., Chartkunchand, K.~C., {et~al.} 2017, Phys. Rev.
  Lett., 119, 073001, \dodoi{10.1103/PhysRevLett.119.073001}

\bibitem[{Seitz {et~al.}(2013)Seitz, Zettergren, Rousseau, Wang, Chen,
  Gatchell, Alexander, Stockett, Rangama, Chesnel, Capron, Poully, Domaracka,
  M{\'e}ry, Maclot, Vizcaino, Schmidt, Adoui, Alcam{\'\i}, Tielens,
  Mart{\'\i}n, Huber, \& Cederquist}]{Seitz:2013wc}
Seitz, F., Zettergren, H., Rousseau, P., {et~al.} 2013, The Journal of Chemical
  Physics, 139, 034309, \dodoi{10.1063/1.4812790}

\bibitem[{Sellgren {et~al.}(2010)Sellgren, Werner, Ingalls, Smith, Carleton, \&
  Joblin}]{Sellgren:2010aa}
Sellgren, K., Werner, M.~W., Ingalls, J.~G., {et~al.} 2010, The Astrophysical
  Journal Letters, 722, L54.
\newblock \url{http://stacks.iop.org/2041-8205/722/i=1/a=L54}

\bibitem[{Slavin {et~al.}(2004)Slavin, Jones, \& Tielens}]{Slavin:2004aa}
Slavin, J.~D., Jones, A.~P., \& Tielens, A. G. G.~M. 2004, The Astrophysical
  Journal, 614, 796, \dodoi{10.1086/423834}

\bibitem[{St{\o}chkel \& Andersen(2013)}]{Stochkel:2013aa}
St{\o}chkel, K., \& Andersen, J.~U. 2013, The Journal of Chemical Physics, 139,
  164304, \dodoi{10.1063/1.4826097}

\bibitem[{Stockett {et~al.}(2018)Stockett, Wolf, Gatchell, Schmidt, Zettergren,
  \& Cederquist}]{Stockett:2018wq}
Stockett, M.~H., Wolf, M., Gatchell, M., {et~al.} 2018, Carbon, 139, 906 ,
  \dodoi{10.1016/j.carbon.2018.07.073}

\bibitem[{Thomas {et~al.}(2011)Thomas, Schmidt, Andler, Bj{\"o}rkhage, Blom,
  Br{\"a}nnholm, B{\"a}ckstr{\"o}m, Danared, Das, Haag, Halld{\'e}n, Hellberg,
  Holm, Johansson, K{\"a}llberg, K{\"a}llersj{\"o}, Larsson, Leontein, Liljeby,
  L{\"o}fgren, Malm, Mannervik, Masuda, Misra, Orb{\'a}n, Pa{\'a}l, Reinhed,
  Rensfelt, Ros{\'e}n, Schmidt, Seitz, Simonsson, Weimer, Zettergren, \&
  Cederquist}]{Thomas:2011aa}
Thomas, R.~D., Schmidt, H.~T., Andler, G., {et~al.} 2011, Review of Scientific
  Instruments, 82, 065112, \dodoi{10.1063/1.3602928}

\bibitem[{Tielens(2008)}]{Tielens:2008aa}
Tielens, A. G. G.~M. 2008, Annual Review of Astronomy and Astrophysics, 46,
  289, \dodoi{10.1146/annurev.astro.46.060407.145211}

\bibitem[{Tielens(2013)}]{Tielens:2013aa}
---. 2013, Reviews of Modern Physics, 85, 1021,
  \dodoi{10.1103/RevModPhys.85.1021}

\bibitem[{Tomita {et~al.}(2001{\natexlab{a}})Tomita, Andersen, Gottrup,
  Hvelplund, \& Pedersen}]{Tomita:2001ab}
Tomita, S., Andersen, J.~U., Gottrup, C., Hvelplund, P., \& Pedersen, U.~V.
  2001{\natexlab{a}}, Physical Review Letters, 87, 073401,
  \dodoi{10.1103/PhysRevLett.87.073401}

\bibitem[{Tomita {et~al.}(2001{\natexlab{b}})Tomita, Andersen, Gottrup,
  Hvelplund, \& Pedersen}]{Tomita:2001aa}
---. 2001{\natexlab{b}}, Physical Review Letters, 87, 073401,
  \dodoi{10.1103/PhysRevLett.87.073401}

\bibitem[{Tomita {et~al.}(2002)Tomita, Hvelplund, Nielsen, \&
  Muramoto}]{Tomita:2002aa}
Tomita, S., Hvelplund, P., Nielsen, S.~B., \& Muramoto, T. 2002, Physical
  Review A, 65, 043201, \dodoi{10.1103/PhysRevA.65.043201}

\bibitem[{Wang {et~al.}(2014)Wang, Zettergren, Rousseau, Chen, Gatchell,
  Stockett, Domaracka, Adoui, Huber, Cederquist, Alcam\'{i}, \&
  Mart\'{i}n}]{PhysRevA.89.062708}
Wang, Y., Zettergren, H., Rousseau, P., {et~al.} 2014, Physical Review A, 89,
  062708, \dodoi{10.1103/PhysRevA.89.062708}

\bibitem[{Webster(1992)}]{Webster:1992aa}
Webster, A. 1992, Monthly Notices of the Royal Astronomical Society, 257, 463,
  \dodoi{10.1093/mnras/257.3.463}

\bibitem[{Yamada {et~al.}(2018)Yamada, Chiba, Hirano, \&
  Saitoh}]{Yamada:2018aa}
Yamada, K., Chiba, A., Hirano, Y., \& Saitoh, Y. 2018, AIP Conference
  Proceedings, 2011, 050020, \dodoi{10.1063/1.5053318}

\bibitem[{Zettergren {et~al.}(2010)Zettergren, Johansson, Schmidt, Jensen,
  Hvelplund, Tomita, Wang, Mart{\'\i}n, Alcam{\'\i}, Manil, Maunoury, Huber, \&
  Cederquist}]{Zettergren:2010aa}
Zettergren, H., Johansson, H. A.~B., Schmidt, H.~T., {et~al.} 2010, The Journal
  of Chemical Physics, 133, 104301, \dodoi{http://dx.doi.org/10.1063/1.3479584}

\bibitem[{Zettergren {et~al.}(2013)Zettergren, Rousseau, Wang, Seitz, Chen,
  Gatchell, Alexander, Stockett, Rangama, Chesnel, Capron, Poully, Domaracka,
  M\'ery, Maclot, Schmidt, Adoui, Alcam\'{i}, Tielens, Mart\'{i}n, Huber, \&
  Cederquist}]{Zettergren:2013vp}
Zettergren, H., Rousseau, P., Wang, Y., {et~al.} 2013, Physical Review Letters,
  110, 185501, \dodoi{10.1103/PhysRevLett.110.185501}

\bibitem[{Zhang {et~al.}(2013)Zhang, Bowles, Bearden, Ray, Fuhrer, Ye, Dixon,
  Harich, Helm, Olmstead, Balch, \& Dorn}]{Zhang:2013aa}
Zhang, J., Bowles, F.~L., Bearden, D.~W., {et~al.} 2013, Nat Chem, 5, 880.
\newblock \url{http://dx.doi.org/10.1038/nchem.1748}

\bibitem[{Zhang {et~al.}(2020)Zhang, Sadjadi, \& Hsia}]{Zhang:2020aa}
Zhang, Y., Sadjadi, S., \& Hsia, C.-H. 2020, Astrophysics and Space Science,
  365, 67, \dodoi{10.1007/s10509-020-03779-5}

\bibitem[{Zhen {et~al.}(2014)Zhen, Castellanos, Paardekooper, Linnartz, \&
  Tielens}]{Zhen:2014aa}
Zhen, J., Castellanos, P., Paardekooper, D.~M., Linnartz, H., \& Tielens, A. G.
  G.~M. 2014, The Astrophysical Journal Letters, 797, L30,
  \dodoi{10.1088/2041-8205/797/2/L30}

\bibitem[{Ziegler {et~al.}(1985)Ziegler, Biersack, \& Littmark}]{zbl_pot_book}
Ziegler, J.~F., Biersack, J.~P., \& Littmark, U. 1985, In The Stopping and
  Range of Ions in Matter (New York: Pergamon)

\end{thebibliography}
\bibliographystyle{aasjournal}

\end{document}